\documentclass[aps,prl,amsmath,amssymb,superscriptaddress,twocolumn]{revtex4}
\usepackage{graphicx}

\begin{document}

\title{Anisotropic effect of a magnetic field on the neutron spin resonance in FeSe}

\author{Tong Chen}
\affiliation{Department of Physics and Astronomy, Rice University, Houston, Texas 77005, USA}
\author{Youzhe Chen}
\affiliation{Department of Physics and Astronomy, Johns Hopkins University, Baltimore, Maryland 21218, USA}
\author{David W. Tam}
\affiliation{Department of Physics and Astronomy, Rice University, Houston, Texas 77005, USA}
\author{Bin Gao}
\affiliation{Department of Physics and Astronomy, Rice University, Houston, Texas 77005, USA}
\author{Yiming Qiu}
\affiliation{NIST Center for Neutron Research, National Institute of Standards and Technology, Gaithersburg, Maryland 20899, USA}
\author{Astrid Schneidewind}
\affiliation{J\"{u}lich Center for Neutron Sciences, Forschungszentrum J\"{u}lich GmbH, Outstation at MLZ, D-85747 Garching, Germany}
\author{Igor Radelytskyi}
\affiliation{J\"{u}lich Center for Neutron Sciences, Forschungszentrum J\"{u}lich GmbH, Outstation at MLZ, D-85747 Garching, Germany}
\author{Karel Prokes}
\affiliation{Helmholtz Zentrum Berlin f\"{u}r Materialien und Energie, 14109 Berlin, Germany}
\author{Songxue Chi}
\affiliation{Neutron Scattering Division, Oak Ridge National Laboratory, Oak Ridge, Tennessee 37831, USA}
\author{Masaaki Matsuda}
\affiliation{Neutron Scattering Division, Oak Ridge National Laboratory, Oak Ridge, Tennessee 37831, USA}
\author{Collin Broholm}
\affiliation{Department of Physics and Astronomy, Johns Hopkins University, Baltimore, Maryland 21218, USA}
\affiliation{NIST Center for Neutron Research, National Institute of Standards and Technology, Gaithersburg, Maryland 20899, USA}
\author{Pengcheng Dai}
\email{pdai@rice.edu}
\affiliation{Department of Physics and Astronomy, Rice University, Houston, Texas 77005, USA}

%\thanks{}


\date{\today}

\begin{abstract}
We use inelastic neutron scattering to study the effect of a magnetic field on the
neutron spin resonance ($E_r=3.6$ meV) of superconducting FeSe ($T_c=9$ K).
While a field aligned along the in-plane direction broadens and suppresses the resonance, 
a $c$-axis aligned field does so much more efficiently, 
consistent with the anisotropic field-induced suppression
of the superfluid density from the heat capacity measurements. These results suggest 
that the resonance in FeSe is associated with the superconducting electrons arising 
from orbital selective quasi-particle excitations between the hole and electron Fermi surfaces.
\end{abstract}

%\keywords{}

\maketitle


Conventional Bardeen-Cooper-Schrieffer (BCS) 
superconductivity in materials such as Aluminum and Tin emerges from the
 pairing of electrons through phonon-mediated attractions and is 
associated with the opening of an isotropic superconducting gap in reciprocal space below $T_c$ \cite{BCS}.  Although there is no consensus for a microscopic theory, high-transition-temperature (high-$T_c$) superconductivity in 
copper- and iron-based materials, derived from 
their antiferromagnetic (AF) ordered parent compounds \cite{norman11,keimer15}, 
 is believed to arise from interactions between itinerant electrons mediated by   
spin fluctuations \cite{scalapino}. One of the key signatures 
is the appearance of a neutron spin resonance mode,
a collective spin excitation with an intensity 
tracking the superconducting order parameter below $T_c$ \cite{scalapino,eschrig06,dai}.  The energy of the resonance, $E_r$, in different superconductors is proportional to either $T_c$ or superconducting gap amplitude \cite{wilson,Inosov2011,GYu2009}. 

In weak-coupling itinerant electron picture, 
the resonance is a bound state (spin exciton) appearing below the particle-hole continuum
at a momentum transfer $\bf Q$ that connects parts of the Fermi surface exhibiting a sign-change in the superconducting order parameter \cite{eschrig06}. 
For copper oxide superconductors, which are single band superconductors with $d$-wave gap symmetry \cite{Harlingen,Tsuei,Schmid2010}, the resonance peaks at the in-plane AF wavevector ${\bf Q}_{\rm AF} = (0.5,0.5)$ and displays hourglass-like dispersion around ${\bf Q}_{\rm AF}$ 
consistent with expectations of the spin-exciton picture \cite{Bourges00,Dai01,Reznik04,Hayden04}.
In the absence of (or for very weak) spin-orbit coupling (SOC) \cite{Headings11}, the resonance is isotropic 
in spin space and arises from the spin-1 singlet-triplet
excitations of the electron Cooper pairs \cite{eschrig06,Lipscombe10}.  
When a magnetic field is applied, the spin-1 of the
resonance should split into three energy levels following the Zeeman
energy $E_{\pm}=E_r \pm g \mu_B B$ (at energies $E_r-g\mu_B B$, $E_r$, and 
$E_r+g\mu_B B$) [Fig. 1(a)], where $g = 2$ is the Land$\rm \acute{e}$ 
factor, $B$ is the magnitude of the field, and $\mu_B$ is the Bohr magneton. 
On the other hand, if superconductivity coexists with AF order or there is large SOC-induced  anisotropy, the resonance can be a doublet where the application of a magnetic field will
split the mode into two peaks [Fig. 1(b)] \cite{Lipscombe10}, as seen in the heavy Fermion superconductor CeCoIn$_5$\cite{stock12,raymond}. However, the application of a 14 T magnetic 
field approximately along the $c$-axis in cuprate superconductors, where $T_c$ and the superconducting gap is two orders of magnitude larger, only slightly suppresses the 
intensity of the resonance with no evidence for Zeeman splitting \cite{Dai00,Tranquada04}.

In the case of iron-based superconductors \cite{stewart}, where electrons in Fe 3$d$ $t_{2g}$ band with  
$d_{xz}$, $d_{yz}$, and $d_{xy}$ orbitals are near the Fermi level, superconductivity may occur in multiple orbitals through the hole-electron Fermi surface nesting \cite{hirschfeld}.
As a consequence, the resonance can have more than one component in energy \cite{CZhang2013,Xie2018} and be anisotropic in spin space due to SOC \cite{PSteffens2013,Luo2013,Song2017}. 
Since the effect of Zeeman energy for a maximum possible applied field of 14 T
is still small compared with the intrinsic energy width of the resonance for optimally doped 
iron pnictide/chalcogenide superconductors \cite{Qiu2009,Zhao2010}, 
there is no confirmed evidence of Zeeman field-induced triplet splitting of the resonance \cite{shiliang11,jinsheng10,yu18,juanjuan18}. 
Nevertheless, a $c$-axis aligned magnetic field suppresses the intensity of the mode much more efficiently than for an in-plane field \cite{Zhao2010,shiliang11}. These results are consistent with lower upper critical fields required to suppress superconductivity in $c$-axis aligned fields \cite{stewart}, 
 suggesting that the intensity of the resonance is a measure of superconducting electron pairing density \cite{MWang2013}.

To further test if the resonance in iron-based superconductors is a spin exciton and associated with singlet-triplet or singlet-doublet transition [Figs. 1(a) and 1(b)], we need to find
a clean material with relatively low $T_c$ and a sharp resonance in energy without multi-orbital effects.
FeSe, which undergoes a tetragonal-to-orthorhombic structural transition at $T_s=90$ K, forms a nematic phase below $T_s$, and becomes superconducting at $T_c=9$ K \cite{Hsu08,mar08,mcqueen09,bohmer17}, is an excellent choice for several reasons [Figs. 1(c) and 1(d)].  
First, the compound is known to be extremely clean and has a relatively low
resonance energy of $E_r=3.6$ meV \cite{wang16}. Second, superconductivity in FeSe is 
orbital selective and occurs mostly through hole-electron Fermi surface nesting of quasi-particles with 
$d_{yz}$ orbital characters \cite{Sprau}, resulting in a resonance only 
at the in-plane AF wave vector ${\bf Q}_{\rm AF} = (1,0)$ \cite{chen19}. Third, neutron polarization analysis
of the resonance reveals that the mode is anisotropic in spin space and essentially 
$c$-axis polarized due to SOC \cite{ma17}, suggesting that a 
magnetic field cannot split the mode into triplets. Finally, the upper critical fields to suppress superconductivity in FeSe are around $16$ T and $28$ T for the $c$-axis and in-plane 
aligned fields, respectively \cite{ma17,audo15}, meaning that an applied magnetic field will have a larger impact on superconductivity compared with that of optimally doped iron pnictides.

We carried out inelastic neutron scattering experiments to study the effect of a magnetic field on the resonance
of FeSe using the multi-axis crystal spectrometer (MACS) at NIST Center For Neutron Research \cite{jose08}, and the cold neutron triple-axis spectrometer PANDA at Heinz Maier-Leibnitz Zentrum, Germany \cite{astrid15}. The $c$-axis aligned magnetic field experiments were performed on MACS with a fixed $E_f=3.7$ meV and PANDA with a 
fixed $E_f=5.1$ meV. The vertical magnetic fields were aligned along the $[0,0,1]$ direction perpendicular to the $[H,K,0]$ scattering plane. The in-plane magnetic field experiment was performed on MACS with the same instrumental setup, while the sample was aligned in the $[H,0,L]$ scattering plane with field along the $[0,1,0]$ direction.  Since an in-plane magnetic field will not produce orbital current within the FeSe layer, its effect on the resonance will be mostly the Zeeman effect. 

\begin{figure}[t]
\includegraphics[scale=.75]{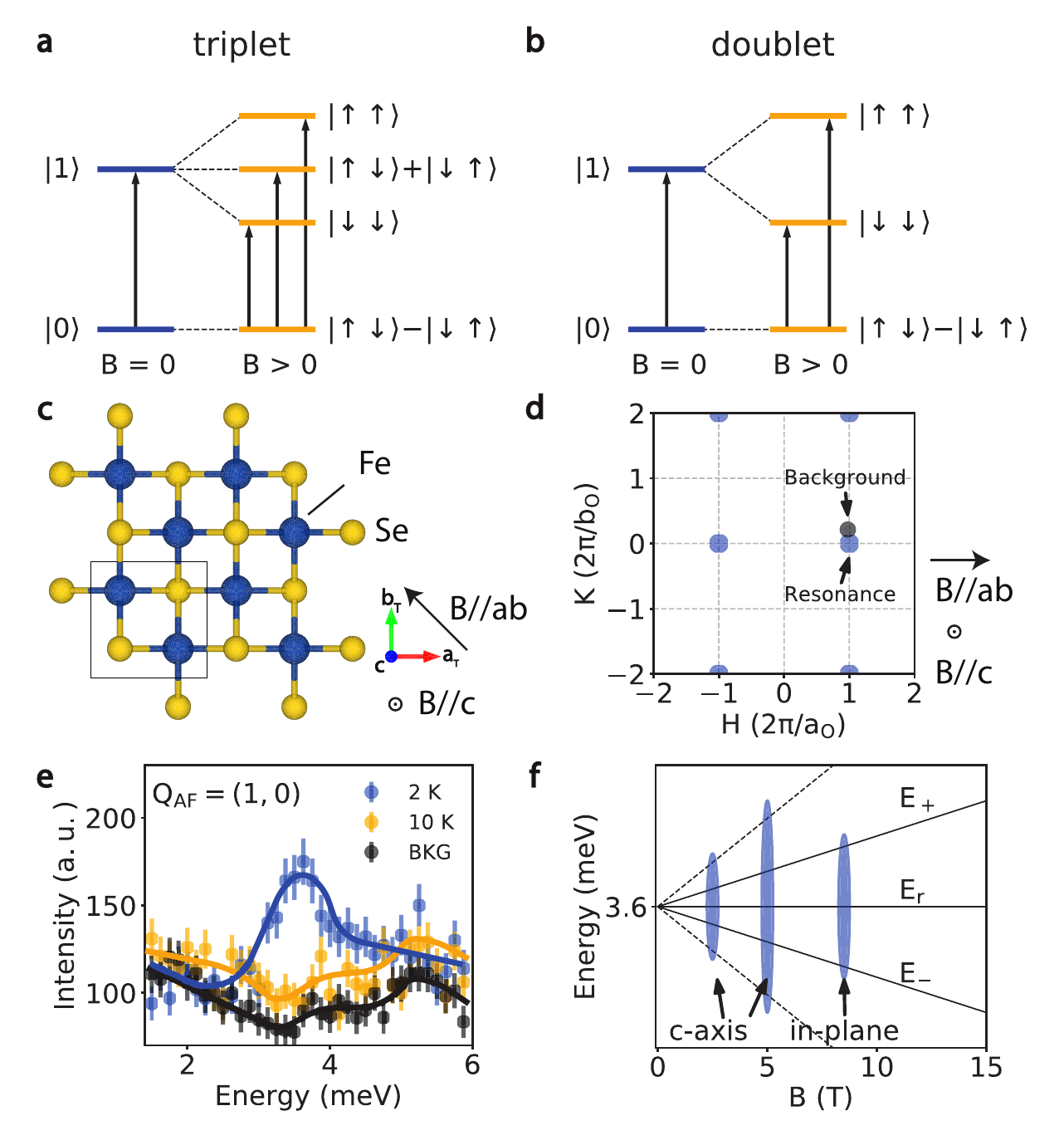}
\caption{
(a) Schematic illustration of the Zeeman splitting of the spin-exciton from singlet $|0\rangle$ to triplet $|1\rangle$ excited states.
(b) Schematic illustration of a singlet-to-doublet excitation.
(c) Crystal structure of FeSe.
(d) Reciprocal space where the blue dots 
represent the ${\bf Q}_{\rm AF} = (1,0)$ wave-vector. The background
position at ${\bf Q}_{\rm bkgd}=(0.977,0.213,0)$ is marked as small circle.
(e) PANDA measurements of the energy dependence of the
 scattering below (blue circles) and above (Yellow circles) 
$T_c$ at ${\bf Q}_{\rm AF} = (1,0)$.  The background scattering is shown
as black circles. The error bars indicate statistical errors of 1 standard deviation.    
(f) Schematic of normalized peaks and excitation positions of the resonance in FeSe  
as a function of increasing magnetic field. 
Solid lines are $E_{\pm}=E_r \pm 2 \mu_B B$ and $E_r$. 
Dashed lines are guides to the eye for a $c$-axis aligned field. \label{Fig1}}
\end{figure}

\begin{figure}[t]
\includegraphics[scale=.75]{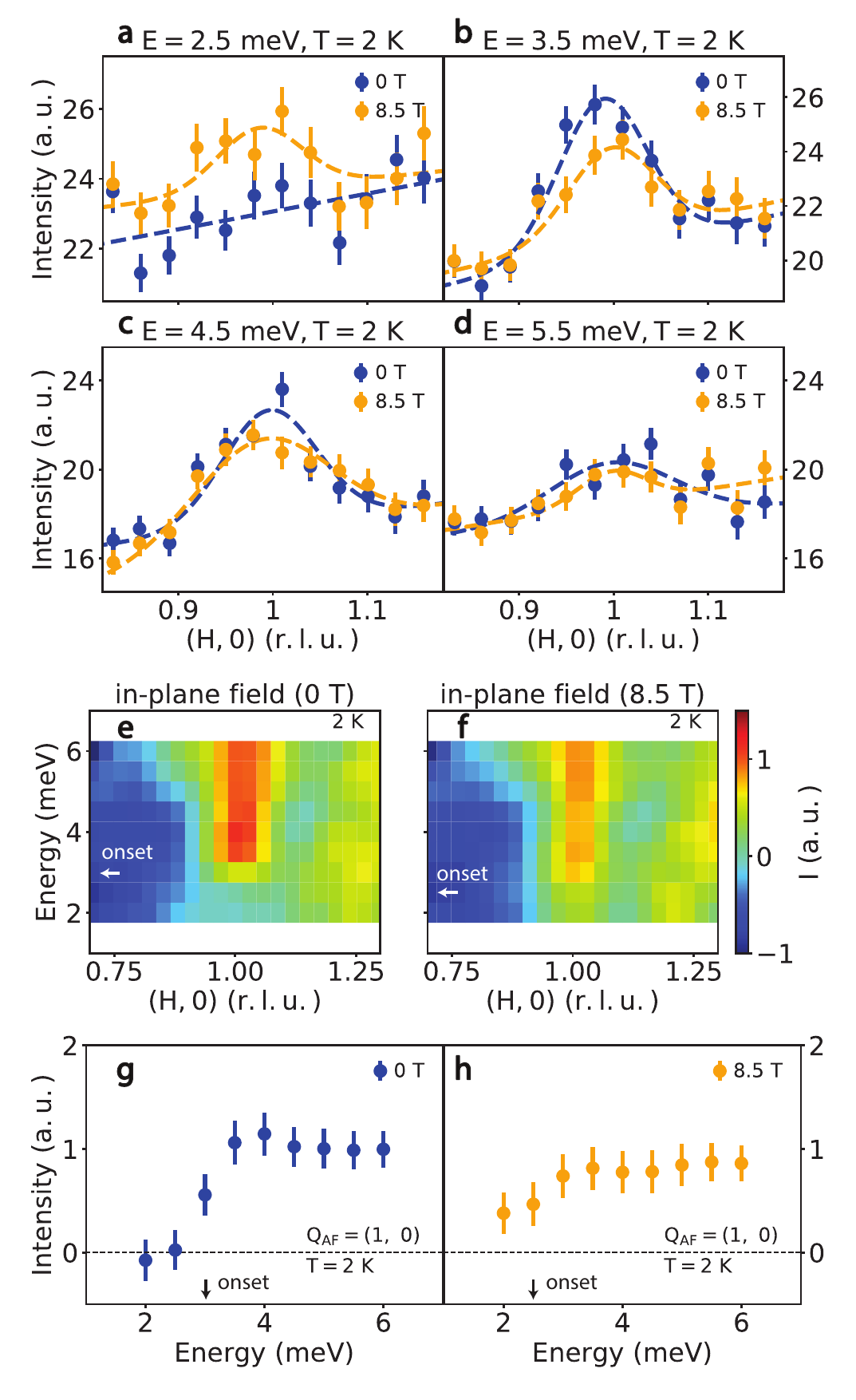}
\caption{
(a)-(d) Constant-energy scans along the $[1,0]$ direction at $E=2.5$, $3.5$, $4.5$ and $5.5$ meV in zero and $8.5$ T in-plane magnetic fields at $T = 2$ K. 
(e) and (f) 2D images of wave-vector and energy dependence of the spin fluctuations in $0$ T and $8.5$ T in-plane magnetic fields at $T = 2$ K.
(g) and (h) are constant-{\bf Q} cuts at the ${\bf Q}_{\rm AF}$ position from (e) and (f), respectively. They have been smoothed two times by the 2D data processing method in David-Mslice program at NCNR. 
The arrows in (e), (f), (g), and (h) indicate the lowest energy where a Gaussian can be fit to the data. The scattering of an assembly of Al plates coated with CYTOP, as well as a constant adjusted to force the scattering at $E=2.4$ meV and ${\bf Q}_{\rm AF}$ to be zero [Fig. 4(b)], was subtracted as background in (e), (f), (g), and (h) \cite{SI}. The monitor counts in (e), (f), (g), 
and (h) are normalized to an arbitrary unit (a.u.) and can be compared directly.
$L$ is integrated in all panels, because spin fluctuations have no $c$-axis modulations in   FeSe. The error bars indicate statistical errors of 1 standard deviation.
\label{Fig2}}
\end{figure}

\begin{figure}[t]
\includegraphics[scale=.75]{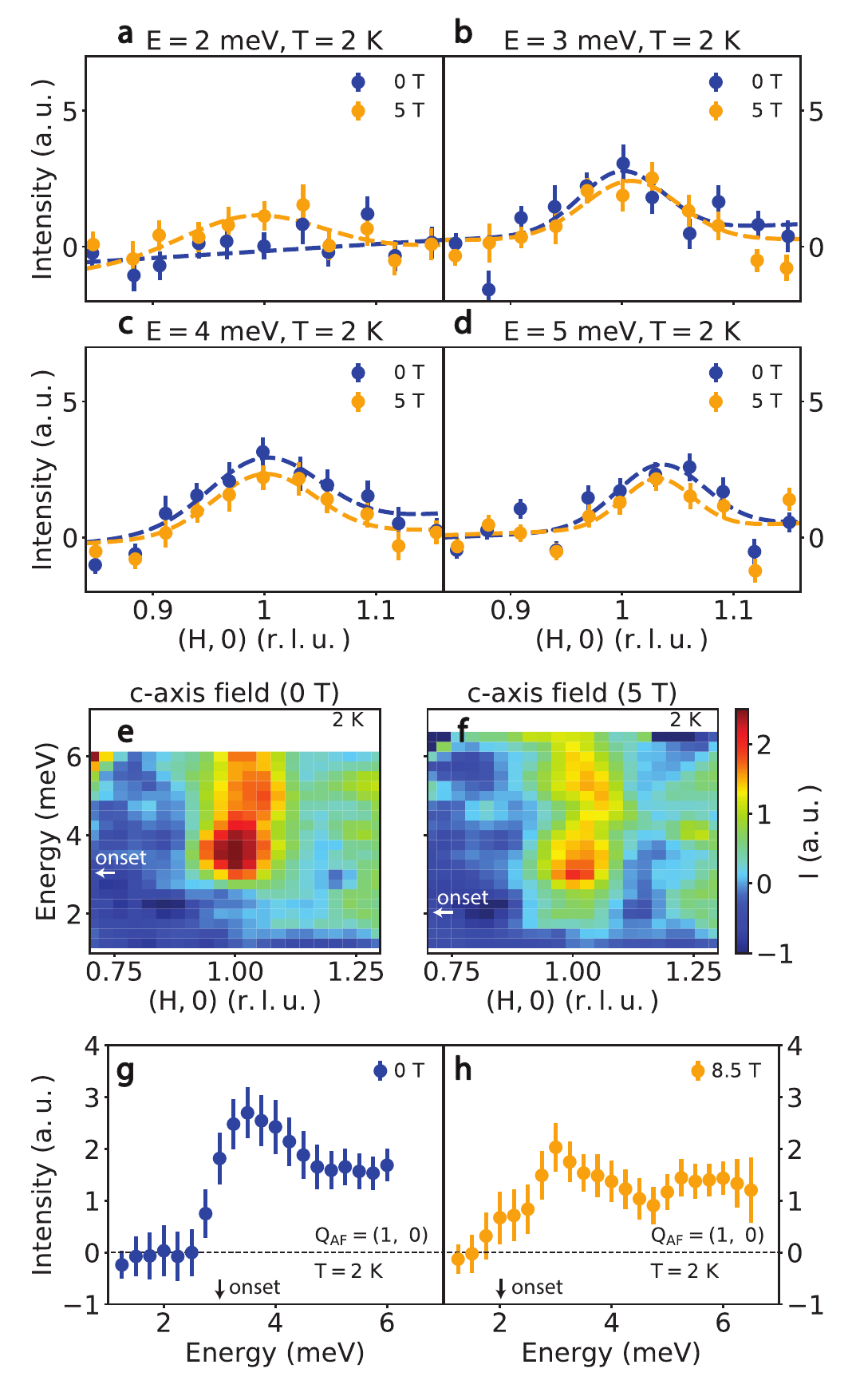}
\caption{
(a)-(d) Constant-energy scans along the $[1,0]$ direction
 at $E=2$, $3$, $4$ and $5$ meV in zero and $5$ T $c$-axis aligned magnetic fields at 
$T=2$ K.  (e) and (f) 2D images of wave-vector and energy dependence of the resonance in zero and $5$ T magnetic fields at 2 K.  The background subtraction process is similar to that
of Fig. 2 \cite{SI}. (g) and (h) are constant-{\bf Q} cuts at the ${\bf Q}_{\rm AF}$ position from (e) and (f), respectively. They have been smoothed two times by the 2D data processing method in David-Mslice program at NCNR.
The arrows in (e), (f), (g), and (h) indicate the lowest energy 
where a Gaussian can be fit to the data. 
The error bars indicate statistical errors of 1 standard deviation.\label{Fig3}}
\end{figure}

\begin{figure}[t]
\includegraphics[scale=.75]{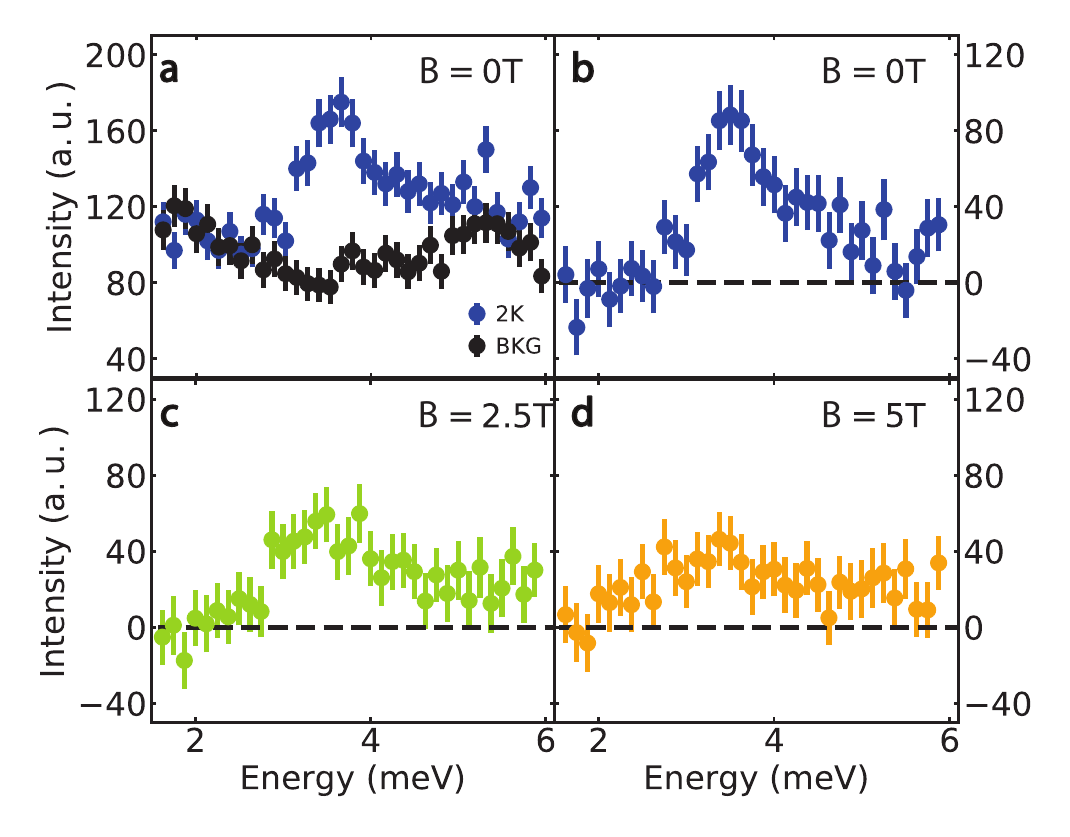}
\caption{
(a) Constant-{\bf Q} scans at ${\bf Q}_{\rm AF}$ and the off-peak background positions   
 at 2 K in zero field as shown in Fig. 1(d).
(b)-(d) Constant-{\bf Q} scans with background subtracted in $0$, $2.5$ and $5$ T $c$-axis aligned fields.
The error bars indicate statistical errors of 1 standard deviation.\label{Fig4}}
\end{figure}

At zero field, superconductivity in FeSe induces a resonance at 
$E_r\approx 3.6$ meV and a spin gap of about $2.8$ meV as shown from data obtained on PANDA [Figs. 1(e) and 1(f)] \cite{wang16,chen19,ma17}.  The effect of an $8.5$ T in-plane magnetic field on the resonance and low-energy spin excitations 
is shown using data obtained on MACS. 
Figures 2(a-d) show constant-energy scans along the $[1,0]$ direction at different energies 
with $8.5$ T and zero magnetic fields in the superconducting state ($T=2$ K).
At $E=2.5$ meV, an $8.5$ T field induces magnetic scattering near ${\bf Q}_{\rm AF} = (1,0)$ 
above the flat background, indicating a reduction in spin gap energy [Fig. 2(a)].  Near the resonance around 
$E=3.5$ [Fig. 2(b)], the field suppresses the resonance as expected.  Above the resonance 
energy at $E=4.5$ and $5.5$ meV, the applied field has little effect on the magnetic scattering [Figs. 2(c), and 2(d)]. Figures 2(e) and 2(f) show the two-dimensional (2D) wave vector-energy images of the resonance above background scattering at zero and    
$8.5$ T field, respectively \cite{SI}.  The effect of an $8.5$ T in-plane magnetic field is to weaken and broaden
the resonance, with no convincing evidence for the splitting of the mode. 
Figure 2(g) is a cut along the energy direction at ${\bf Q}_{\rm AF} = (1,0)$, which reveals
the resonance at 0 T field.  The net effect of a magnetic field is to push the 
spectral weight of the resonance to lower energies [Fig. 2(h)].

Figure 3 illustrates the effect of a $5$ T $c$-axis aligned magnetic field on the resonance. Figure 3(a-d) show constant-energy scans along the $[1,0]$ direction with different energies in $5$ T and zero magnetic fields in the superconducting state ($T=2$ K). At $E=2$ meV, a $5$ T $c$-axis aligned field induces magnetic scattering near ${\bf Q}_{\rm AF}$, which is $1.6$ meV below the spin resonance energy $E_r$. Off the resonance energy at $E=3$ and $4$ meV and above the resonance energy at $E=5$ meV, the applied field has slight effect on the resonance.
Figures 3(e) and 3(f) compare the 2D images of the wave-vector and energy dependence of the spin resonance in $0$ and $5$ T, respectively. For a $c$-axis aligned magnetic field, the upper critical field $B_{c2}(\perp)$ is around $16$ T, meaning that a $5$ T field is already
$\sim$31\% of $B_{c2}$, which is similar to the fraction of 30\% achieved for the 8.5 T in-plane oriented field given the 28 T critical field. Although qualitatively the broadening in energy is similar to 
that of the in-plane field, the amplitude of the broadening is more significant. 
Figures 3(g) and 3(h) show the constant-${\bf Q}$ cuts 
at the ${\bf Q}_{\rm AF}$ position from (e) and (f), respectively.  We see that a applied field shifts the magnetic spectral weight to lower energies.
By comparing Figs. 2(g), 2(f), 3(g) and 3(f), we conclude that a 5 T $c$-axis aligned field has a larger impact on the resonance than that of an $8.5$ T in-plane field.

To determine if the broadening of the resonance in the $c$-axis aligned magnetic field 
follows expectations from the field-induced Zeeman effect,  we carried out additional measurements on PANDA. Figures 4(a) and 4(b) show the evolution of the magnetic 
scattering at ${\bf Q}_{\rm AF} = (1,0)$ in the superconducting state before and after correcting for the background scattering, respectively.
As expected, we see a well-defined spin gap below $2.8$ meV and a resonance peaked
at $E_r=3.6$ meV [Fig. 4(b)].  Upon application of a $2.5$ T field, the resonance broadens and weakens, but still seems to be centered around $E_r=3.6$ meV [Fig. 4(c)]. At $5$ T, the
magnetic scattering is broadened and weakened further [Fig. 4(d)].

To understand these results, we first consider the effect of Zeeman energy $\pm g \mu_B B$
on the resonance. For an isotropic resonance with weak SOC, such as for cuprate superconductors \cite{Headings11}, the application of a magnetic field is expected to split the mode
into a triplet \cite{ismer07}.  This is similar to the magnetic field effect on quantum magnets such as TlCuCl$_3$ \cite{ruegg} and Sr$_{14}$Cu$_{24}$O$_{41}$ \cite{Lorenzo07}, in which the ground state is a singlet and the excited state is a triplet and the system undergoes a so-called magnon Bose-Einstein-Condensation (BEC) 
in magnetic fields \cite{gia08}. When SOC becomes important, as in the case of iron-based 
superconductors \cite{Borisenko16}, low-energy spin excitations become anisotropic in
spin space \cite{Scherer18}. In the case of FeSe, neutron polarization analysis suggests 
that the resonance is highly anisotropic in spin space \cite{ma17}. 
As a consequence, the resonance should not be split by a Zeeman field into a triplet.
If the resonance is a magnon-like excitation, two polarizations perpendicular to the applied field 
are needed to form a doublet. Since the resonance 
is reported to be mostly polarized along the $c$-axis \cite{ma17}, a Zeeman field should 
be unable to split the mode into a doublet. 

Figure 1(f) compares the expected broadening of the resonance assuming that the mode splits into three peaks in the applied magnetic fields via the Zeeman effect. 
Taking $g = 2$, the field-induced Zeeman splitting equals to $0.58$ and $0.98$ meV in $5$ and $8.5$ T, respectively. In the $8.5$ T in-plane magnetic field, the lowest energy where excitation can be observed at ${\bf Q}_{\rm AF}$ is $2.5$ meV, which is $1.1$ meV below the peak of the resonance at a zero field. For a $5$ T field along the $c$-axis, the magnetic signal can be observed down to $2$ meV.  Since the Zeeman splitting should have no field directional dependence, the wider in-plane field-induced resonance must be due to field-induced orbital current that suppresses superconductivity.  As a function of the increasing
magnetic field along the $c$-axis, the intensity of the resonance is gradually suppressed and broadened, qualitatively consistent with the field-induced suppression of superconductivity and super-fluid density \cite{kasa14}.  Indeed, as previously noted the experiments correspond to applying the essentially the same 30\% fraction of the upper critical field for both field directions.

In recent electric and thermal transport measurements \cite{kasa14}, it was argued that FeSe is in a BCS-BEC cross over regime, and a large magnetic field along the $c$-axis might induce a new superconducting phase coexisting with magnetic order, possibly the FFLO state \cite{kasa14,chen17,shi18,Kasahara19}.  
To study if this phase has field-induced magnetic order as suggested from the field-induced broadening of the resonance, we carried out neutron diffraction experiments using 
 the 2-axis-diffractometer E4, HZB, Germany \cite{prokes17}. 
 We aligned about ~200 pieces of FeSe single crystals in the $[H,K,0]$ scattering plane and mapped out one quadrant of the zone with wave-vector between $0.14$ and $1.54$ reciprocal lattice unit \cite{SI}. However, we did not find any observable difference between data at different temperatures ($0.25\ K$ and $3\ K$) or at base temperature ($0.25\ K$) with different fields ($0$, $12$, $14$, and $14.5$-T) along the $c$-axis, suggesting no observable 
field-induced magnetic order up to $14.5$ T \cite{SI}.  However,  
thermal conductivity data indicated the FFLO phase might exist for a 
$\sim$24 T in-plane magnetic field \cite{Kasahara19}. Unfortunately, currently available neutron spectrometers cannot access such a high DC field..

Assuming that the resonance is directly associated with superconducting 
electron pairs \cite{MWang2013}, we can estimate the upper critical fields for $c$-axis
and in-plane fields using field-induced suppression of the resonance. If 
the spin gap energy below the resonance decreases linearly with the applied field, 
we estimate that the lowest energy position of the spin gap to $E=0$ meV in the $c$-axis and in-plane magnetic fields corresponds to fields of $12$ and $30$ T, respectively.
These values are close to the measured upper critical fields ($B_{c2}$) of $16$ and $28$ T. 
The field directional dependence of the spin resonance is also consistent with that of the superfluid density from heat capacity measurements \cite{lin11}, implying that the resonance is associated with superconducting electrons arising from 
the orbital selective hole-electron quasi-particle excitations \cite{Sprau,chen19}.

In summary, we determined the effect of $c$-axis and in-plane magnetic fields on the 
neutron spin resonance of FeSe. We find that an in-plane magnetic increases the width of the
resonance following the field-induced Zeeman effect. A $c$-axis aligned field 
suppresses and broadens the resonance much more effectively than the in-plane field, 
clearly related to the orbital effect and vortex currents induced by the $c$-axis field. 
The data indicates that rather than the absolute applied field, it is the ratio of the applied field to the upper critical field that controls changes in the magnetic excitation spectrum.
Our results are consistent with the hypothesis that the resonance is associated with electron pairing
density in FeSe superconductor. 

We would like to thank R. Feyerherm from HZB Berlin for set-up and 
operating the dilution refrigerator.
Neutron scattering work at Rice is supported by the US Department of Energy, BES DE-SC0012311 (P.D.). The single-crystal synthesis work at Rice is supported by 
Robert A. Welch Foundation grant no. C-1839 (P.D.). Sample preparation at Johns Hopkins University is supported by the US Department of Energy grant no. DE-SC0019331. The access to MACS was provided by the Center for High Resolution Neutron Scattering, a partnership between the National Institute of Standards and Technology and the National Science Foundation under agreement No. DMR-1508249.

\end{document}